\documentclass[12pt,a4paper]{article}
\usepackage[latin1]{inputenc}
\usepackage{amsmath}
\usepackage{amsfonts}
\usepackage{amssymb,graphicx,cite}
\newcommand{\be}{\begin{equation}}
\newcommand{\ee}{\end{equation}}
\newcommand{\ba}{\begin{eqnarray}}
\newcommand{\ea}{\end{eqnarray}}
\newcommand{\nl}{\nonumber \\}

\begin{document}


\title{Towards understanding broad degeneracy in non-strange mesons}

\author{S.S.~Afonin}
\maketitle
\begin{center}
\it V.A. Fock Institute of Physics,
St. Petersburg State University, \\
St. Petersburg 198504, ul. Ulyanovskaya 1, Russia\\
E-mail: afonin24@mail.ru
\end{center}

\begin{abstract}
The spectroscopic regularities of modern empirical data on the
non-strange mesons up to 2.4~GeV can be summarized as a systematic
clustering of states near certain values of energy. It is getting
evident that some unknown X-symmetry triggers the phenomenon.
We review the experimental status of this symmetry and recent
theoretical attempts put forward for explanation of broad
degeneracy.
\end{abstract}



\thispagestyle{empty}

\newpage


\section{Introduction}

The spectroscopic studies of light hadrons have a long history reflected in the numerous
literature.
In recent years many new data has appeared in the section "Further States" of
Particle Data~\cite{pdg}. The emerging spectroscopic picture is quite intriguing and
theoretical explanations are called for. This raised a renewed interest to the subject.

As was realized recently~\cite{a1,a11,a2,sh,glozman} the experimental spectrum of light non-strange
mesons reveals a broader degeneracy than one could expect from the approximate
symmetries of QCD Lagrangian.
For instance, the experimental spectrum points out an intriguing pattern
of degeneracy among the meson states with different spin, see Fig.~1
below. It is getting evident that some symmetry of unknown nature
triggers the phenomenon, we will refer to it as X-symmetry.

It is quite usual in physics that all spectral symmetries of compound systems are not directly
seen on the fundamental level. In the same way, it may be that the X-symmetry seen in Fig.~1
hardly can be envisaged directly from QCD, rather one should look into how the states are
"constructed".

A popular assumption about the
structure of mesons is the hypothesis that the mesons represent a
gluon string with a quark/antiquark at the ends. We will review
the typical predictions of the hadron string models and the linearly rising
potentials and compare them with the available experimental data, of our concern
will be the spectroscopic aspect of the problem.

Typically the hadron strings and linearly rising potentials predict the following law for the
spectrum of excitations,
\be
\label{1}
M^2\sim a(L+bn),
\ee
where $L$ is the angular momentum, $n$ is the radial quantum number, $a$ and $b$ are
constants characterizing the angular and radial slopes, respectively.
As the quark spin is $\frac12$, the quark-antiquark pair can be
either in the singlet ($s=0$) or in the triplet ($s=1$) states.
The angular momentum $L$ is related to the total spin $J$ as
\ba
s=0:\quad J&=&L, \nl
s=1:\quad J&=&L,L\pm1 \quad (\text{for}\, L=0:\,J=L+1).
\label{1.2}
\ea

Recently it was proposed~\cite{sh} that in reality $b=1$. We will
scrutinize this proposal
using the latest experimental data. As a byproduct
we will obtain a form of the string-like spectrum which seems to
be the most consistent with the available phenomenology. This
analysis could be useful for future hadron string approaches,
AdS/QCD methods, and other QCD-inspired models.

The review is organized as follows. In Sect.~2 we consider some
general aspects of quasiclassical string description for the light
mesons. A phenomenological analysis of outlined picture is
performed in Sect~3. The Sect.~4 is devoted to discussions of
obtained results and of some related problems. We conclude in
Sect.~5.

\section{Quasiclassical hadron string picture}

Let us present some heuristic ideas leading to spectrum~\eqref{1}. For high radial or orbital excitation
a meson state can be considered quasiclassically as a pair of relativistic quarks interacting via
a confining linear potential. Consequently, neglecting the quark spin, the meson mass can be written as
\be
\label{2}
M=2p+\sigma r,
\ee
where $p$ is the relativistic quark momentum and $\sigma$ is the mass density per unit length.
The maximal length
of chromoelectric flux tube between the quarks is
\be
\label{3.2}
l=\frac{M}{\sigma}.
\ee
Applying the WKB method (see, {\it e.g.},~\cite{landau}) one obtains
the following quantization condition,
\be
\label{3}
\int\limits_0^lp\,dr=\pi (n+\gamma),\qquad n=0,1,2,\dots
\ee
Substituting the momentum $p$ from Eq.~\eqref{2} one gets
\be
\label{4}
M^2=4\pi\sigma (n+\gamma).
\ee
Here $\gamma$ is a constant of order of unity characterizing the
nature of turning points (in a close vicinity of these points the
semiclassical methods are not applicable). The correction $\gamma$
is known to be of importance for practical applications because it
usually extends the applicability of the Bohr-Sommerfeld
quantization~\eqref{3} from $n\sim10$ to $n\sim1$. In some cases
({\it e.g.}, the harmonic oscillator) the WKB approximation even provides
the exact spectrum.
The analysis performed in~\cite{morgunov} showed that the accuracy
of the WKB method in the given situation is at the level of
several percents. As long as one deals with a
centrosymmetrical potential in Eq.~\eqref{2}, the WKB method gives
\be
\gamma=\frac12.
\ee
This value is quite remarkable: The same value $\frac12$ is predicted
by the Lovelace-Shapiro dual amplitude~\cite{LS,Sh} and
in some channels it appeared naturally within the QCD sum rules~\cite{a3,a31,a32,a33}.
For the S-wave states ($L=0$), however, the situation is different
and the WKB method yields
\be
\gamma_s=\frac34.
\ee

On the other hand, the spectrum of rotating relativistic string is
known to be linear in the angular momentum $L$ (Chew-Frautschi
formula),
\be
\label{cf}
M^2=2\pi\sigma L.
\ee
The first string derivation of Eq.~\eqref{cf} was proposed by
Nambu~\cite{nambu}. The idea is as follows. Suppose that massless
quarks rotate at the speed of light at radius $l/2$. At the
distance $r$ from the center of rotation the speed of flux tube
connecting the quarks is $v(r)=2r/l$. Then the mass of rotating
gluon flux tube is
\be
\label{ft1}
M=2\int\limits_0^{l/2}\frac{\sigma\,dr}{\sqrt{1-v^2(r)}}=\frac{\pi\sigma
l}{2},
\ee
while the angular momentum is
\be
\label{ft2}
L=2\int\limits_0^{l/2}\frac{\sigma r v(r)\,dr}{\sqrt{1-v^2(r)}}=\frac{\pi\sigma
l^2}{8}.
\ee
Combining Eqs.~\eqref{ft1} and~\eqref{ft2} one arrives at
Eq.~\eqref{cf}.

The next step is to quantize the quasiclassical rotating string.
This is a rather controversial problem in the literature. In
particular, one should reconcile a different factor of
proportionality between the mass $M$ and the length of gluon flux tube $l$ in
Eqs.~\eqref{3.2} and~\eqref{ft1}. Usually one obtains the
following relation (at least for large $L$ and $n$)
\be
\label{mm}
M^2=2\pi\sigma(L+bn+c),
\ee
where the parameters $b$ and $c$ are model dependent. The factor
$b$ lies typically in the interval $1\leq b\leq2$~\cite{string,string1,string2,baker}.
At integer values of $b$, in particular $b=1$ and $b=2$, the
specific spectral degeneracies emerge at different $(L,n)$. In the
case $b=2$ the degeneracy is of the harmonic oscillator type. For
$b=1$ one obtains the Coulomb-like degeneracy, where the energy
levels (masses) depend only on the "principal quantum number" $N$,
\be
\label{pqn}
N=L+n+1.
\ee

Let us discuss briefly the intercept $2\pi\sigma c$. For spinless
constituents the physical meaning of this quantity seems to be the
same as in the nonrelativistic Quantum Mechanics: It is the energy
of zero-point oscillations of quarks inside a meson, which appear
due to the uncertainty relation. In the reality, however, quarks do possess the
spin. The resulting spin-spin interaction could give a substantial
contribution to the mass of the ground S-wave mesons, where the
quark/antiquark spin wave function have a maximal region of
overlap. Presently, there is no complete understanding of underlying dynamics.

Finally, some comments are in order. In considering a spectrum like
in Eq.~\eqref{1} (with Eq.~\eqref{1.2})
the following objection appears immediately:
In relativistic systems, the internal angular momentum
and spin cannot be separated, so why we should trust in this model? However,
there are physical arguments that the spin-orbital correlations
are asymptotically suppressed in excited hadrons~\cite{sh,glozman,wil,wil1,wil2,gloz2}.
In this case the usual quantum-mechanical rules for classifying the composite states
seem to be applicable.
In fact, it has been known for long ago that the nonrelativistic classification of light mesons
works rather well~\cite{godfrey,godfrey1}. What is important for us is that this classification
predicts the doubling of states in the channels where the resonances can be created by different
angular momentum. For instance, the vector mesons can have $L=0$ and $L=2$
(the so-called S- and D-wave mesons in the nonrelativistic spectroscopy),
hence, they are doubled. Experimentally such a doubling is well seen. In practice one
achieves the separation of resonances into
the states with different angular momentum by using the polarization data. In particular,
a good separation was obtained for the states with
$(C,I)=(+1,0),(-1,1)$ in the Crystal Barrel experiment~\cite{bugg,bugg1}.
For other channels with duplication of states the experimental separation is not so clear.
We note also that the spectrum under consideration can be cast
into a relativistic form due to Eq.~\eqref{1.2},
\be
\label{mj}
M^2(J,n)\sim J+n+c.
\ee
In this case, however, the constant $c$ is not approximately universal for all channels.

Another objection comes from the lattice calculations which typically point out a string
breaking in the static potential between two quarks at distances of about 1--1.5~fm.
In the potential models one usually mimics the phenomenon by a screened linear potential
(see, {\it e.g.},~\cite{gon}). In this respect we remind that
the QCD forces can be modeled by the static potential only for heavy enough quarks.
The extrapolation of
these results to the light quark sector, generally speaking, is not justified since the
nonstatic contributions, say, the velocity-dependent terms, are expected to be important
for the ultrarelativistic quarks, if not decisive.

\section{Phenomenological analysis}

In this section we will estimate the constants in Eq.~\eqref{mm} from the phenomenology.
The procedure would be quite straightforward if many reliable experimental data
were available. The Particle Data~\cite{pdg} contains a number of reliable non-strange
mesons below
1.9~GeV. Above that region the experimental situation is worse. The only experiment
which systematically looked for the non-strange resonances above 1.9~GeV (in the region
1.9--2.4~GeV) was the one carried out by the Crystal Barrel Collaboration on the
proton-antiproton annihilation in flight,
the corresponding results are summarized in a review~\cite{bugg,bugg1}. The Particle Data
cites these results in the section "Further States".

As a basis we take the analysis performed in~\cite{a2}, where the spectroscopy of non-strange
mesons from the Particle Data (below 1.9~GeV) and Crystal Barrel (in the range 1.9--2.4 GeV)
was analysed. For completeness we add the one-star states from~\cite{bugg} and also the
$\rho(1900)$ and $h_1(1595)$ mesons. We exclude the exotic $\pi_1$-states. Three such mesons
are known with the masses $1376\pm17$, $1653^{+18}_{-15}$, and $2013\pm25$ MeV.
Although the masses of these states agree nicely with the cluster structure of
meson mass distribution, a definite $(L,n)$-assignment cannot be done for them
within the standard Quark Model. Finally, the states we use in the present analysis
are given in Table~1.

\begin{table}
\caption{\small The masses (in MeV) of states from~\cite{pdg} and~\cite{bugg}.
Experimental errors are indicated. Each column corresponds to a cluster in Fig.~1.}
\vspace{0.2cm}
\begin{tabular}{llllll}
\hline
\hline
\noalign{\smallskip}
$\pi$&$135$&
$1300\pm100$&$1812\pm14$& $2070\pm35$& $2360\pm25$\\
$\eta$& 
&$1294\pm4$& $1760\pm11$& $2010^{+35}_{-60}$& $2285\pm20$\\
$\omega$& $782.65\pm0.12$&
$1400\div1450$& $1670\pm30$& $1960\pm25$&\!\!\!\!\!\!
\begin{tabular}{l}
$2205\pm30$\\
$2295\pm50$\\
\end{tabular}\\
$\rho$&$775.5\pm0.4$&
$1459\pm11$& $1720\pm20$& \!\!\!\!\!\!
\begin{tabular}{l}
$1900\pm?$\\
$2000\pm30$\\
\end{tabular}&\!\!\!\!\!\!
\begin{tabular}{l}
$2110\pm35$\\
$2265\pm40$\\
\end{tabular}\\
$f_0$&& 
$1200\div1500$& $1770\pm12$& $2020\pm38$& $2337\pm14$\\
$a_0$&& 
$1474\pm19$&& $2025\pm30$&\\
$a_1$&& $1230\pm40$& $1647\pm22$&$1930^{+30}_{-70}$&$2270^{+55}_{-40}$\\
$f_1$&&  $1281.8\pm0.6$& & $1971\pm15$& $2310\pm60$\\
$h_1$&& $1170\pm20$& $1595\pm20$ & $1965\pm45$& $2215\pm40$\\
$b_1$&& $1229.5\pm3.2$& $1620\pm15$ & $1960\pm35$& $2240\pm35$\\
$f_2$&& $1275.4\pm1.1$& $1638\pm6$& \!\!\!\!\!\!
\begin{tabular}{l}
$1934\pm20$\\
$2001\pm10$\\
\end{tabular}&\!\!\!\!\!\!
\begin{tabular}{l}
$2240\pm15$\\
$2293\pm13$\\
\end{tabular}\\
$a_2$&& $1318.3\pm0.6$& $1732\pm16$&\!\!\!\!\!\!
\begin{tabular}{l}
$1950\pm40$\\
$2030\pm20$\\
\end{tabular}&\!\!\!\!\!\!
\begin{tabular}{l}
$2175\pm40$\\
$2255\pm20$\\
\end{tabular}\\
$\pi_2$&&& $1672.4\pm3.2$& $2005\pm15$& $2245\pm60$\\
$\eta_2$&&& $1617\pm5$& $2030\pm16$& $2267\pm14$\\
$\omega_3$&&& $1667\pm4$& $1945\pm20$& \!\!\!\!\!\!
\begin{tabular}{l}
$2255\pm15$\\
$2285\pm60$\\
\end{tabular}\!\!\!\!\!\!
\\
$\rho_3$&&& $1688.8\pm2.1$& $1982\pm14$&\!\!\!\!\!\!
\begin{tabular}{l}
$2300^{+50}_{-80}$\\
$2260\pm20$\\
\end{tabular}\\
$\omega_2$&&&& $1975\pm20$& $2195\pm30$\\
$\rho_2$&&&& $1940\pm40$& $2225\pm35$\\
$f_3$&&&& $2048\pm8$& $2303\pm15$\\
$a_3$&&&& $2031\pm12$& $2275\pm35$\\
$h_3$&&&& $2025\pm20$& $2275\pm25$\\
$b_3$&&&& $2032\pm12$& $2245\pm50$\\
$a_4$&&&& $2005^{+25}_{-45}$& $2255\pm40$\\
$f_4$&&&& $2018\pm6$& $2283\pm17$\\
$\omega_4$&&&&& $2250\pm30$\\
$\rho_4$&&&&& $2230\pm25$\\
$\pi_4$&&&&& $2250\pm15$\\
$\eta_4$&&&&& $2328\pm38$\\
$\omega_5$&&&&& $2250\pm70$\\
$\rho_5$&&&&& $2300\pm45$\\
\noalign{\smallskip}
\hline
\hline
\end{tabular}
\end{table}

Now we must provide a definite $(L,n)$-assignment for the states in Table~1. For some
mesons the value of angular momentum $L$ is dictated by the quantum numbers of
the quark-antiquark pair. Say, the axial-vector mesons can have only $L=1$, {\it etc}.
For other mesons two different values of $L$ can be assigned, for instance, the vector
mesons can have $L=0$ and $L=2$. The problem appears, how they can be distinguished?
This problem is indeed serious for the energy region above 1.7~GeV, where the doubling
of states with identical quantum numbers in a close mass interval emerges.
As discussed at the end of Sect.~2, for the states
with $(C,I)=(+1,0),\,(-1,1)$ the angular momentum assignment is provided,
in many cases, by the polarisation data~\cite{bugg}. For other states we use a general
principle: The states with a larger $L$ have a larger centrifugal barrier, which pushes
up the corresponding masses. Finally, our assignment in $(L,n)$ is displayed in Table~2.
Reading the data in Table~2 in a diagonal way (for the sake of convenience we
introduced different frames) one can immediately notice an approximate degeneracy
of states with the same $L+n$, just as predicted by Eq.~\eqref{1} if $b=1$.
The data is visualized in Fig.~1, with the experimental errors being indicated.
The clusters of states formed at fixed $L+n$ are distinctively seen.

\begin{table}
\vspace{-1cm}
\caption{
\small Classification of light non-strange mesons according
to the values of $(L,n)$. The states
with the lowest star rating (according to~\cite{bugg}) or which are doubtful
as non-strange quark-antiquark resonances are marked by the question mark.}
\vspace{0.2cm}
\begin{tabular}{cccccc}
\hline
\hline
\begin{tabular}{c}
\begin{picture}(15,15)
\put(0,12){\line(1,-1){15}}
\put(-2,-3){$L$}
\put(10,7){$n$}
\end{picture}\\
\end{tabular}
& 0 & 1 & 2 & 3 & 4 \\
\hline
0
&
\begin{tabular}{c}
$\pi(140)$\\
$\rho(770)$\\
$\omega(780)$\\
\end{tabular}
&
\begin{tabular}{|c|}
\hline
$\pi(1300)$\\
$\rho(1450)$(?)\\
$\omega(1420)$(?)\\
$\eta(1295)$\\
\hline
\end{tabular}
&
\begin{tabular}{c}
$\pi(1800)$\\
$\eta(1760)$\\
\end{tabular}
&
\begin{tabular}{||c||}
\hline
$\pi(2070)$\\
$\rho(1900)$(?)\\
$\eta(2010)$\\
\hline
\end{tabular}
&
\begin{tabular}{c}
$\pi(2360)$\\
$\rho(2150)$\\
$\omega(2205)$(?)\\
$\eta(2285)$\\
\end{tabular}
\\
1
&
\begin{tabular}{|c|}
\hline
$f_0(1370)$\\
$a_0(1450)$(?)\\
$a_1(1260)$\\
$f_1(1285)$\\
$b_1(1230)$\\
$h_1(1170)$\\
$a_2(1320)$\\
$f_2(1275)$\\
\hline
\end{tabular}
&
\begin{tabular}{c}
$f_0(1770)$\\
$a_1(1640)$\\
$b_1(1620)$(?)\\
$h_1(1595)$(?)\\
$a_2(1680)$\\
$f_2(1640)$\\
\end{tabular}
&
\begin{tabular}{||c||}
\hline
$f_0(2020)$\\
$a_0(2025)$\\
$a_1(1930)$(?)\\
$f_1(1971)$\\
$b_1(1960)$\\
$h_1(1965)$\\
$a_2(1950)$(?)\\
$f_2(1934)$\\
\hline
\end{tabular}
&
\begin{tabular}{c}
$f_0(2337)$\\
$a_1(2270)$(?)\\
$f_1(2310)$\\
$b_1(2240)$\\
$h_1(2215)$\\
$a_2(2175)$(?)\\
$f_2(2240)$\\
\end{tabular}
&\\
2
&
\begin{tabular}{c}
$\rho(1700)$\\
$\omega(1650)$\\
$\pi_2(1670)$\\
$\eta_2(1645)$\\
$\rho_3(1690)$\\
$\omega_3(1670)$\\
\end{tabular}
&
\begin{tabular}{||c||}
\hline
$\rho(2000)$\\
$\omega(1960)$\\
$\pi_2(2005)$\\
$\eta_2(2030)$\\
$\rho_2(1940)$\\
$\omega_2(1975)$\\
$\rho_3(1982)$\\
$\omega_3(1945)$\\
\hline
\end{tabular}
&
\begin{tabular}{c}
$\rho(2265)$\\
$\omega(2295)$(?)\\
$\pi_2(2245)$\\
$\eta_2(2267)$\\
$\rho_2(2225)$\\
$\omega_2(2195)$\\
$\rho_3(2300)$(?)\\
$\omega_3(2285)$\\
\end{tabular}
&  &\\
3
&
\begin{tabular}{||c||}
\hline
$f_2(2001)$\\
$a_2(2030)$\\
$f_3(2048)$\\
$a_3(2031)$\\
$b_3(2032)$\\
$h_3(2025)$\\
$f_4(2018)$\\
$a_4(2005)$\\
\hline
\end{tabular}
&
\begin{tabular}{c}
$f_2(2293)$\\
$a_2(2255)$\\
$f_3(2303)$\\
$a_3(2275)$\\
$b_3(2245)$\\
$h_3(2275)$\\
$f_4(2283)$\\
$a_4(2255)$\\
\end{tabular}
&  &  &\\
4
&
\begin{tabular}{c}
$\rho_3(2260)$\\
$\omega_3(2255)$\\
$\rho_4(2230)$\\
$\omega_4(2250)$(?)\\
$\pi_4(2250)$\\
$\eta_4(2328)$\\
$\rho_5(2300)$\\
$\omega_5(2250)$\\
\end{tabular}
&  &  &  &\\
\hline
\hline
\end{tabular}
\end{table}

\begin{center}
\begin{figure}
\vspace{-5cm}
\hspace{-4cm}
\includegraphics[scale=1]{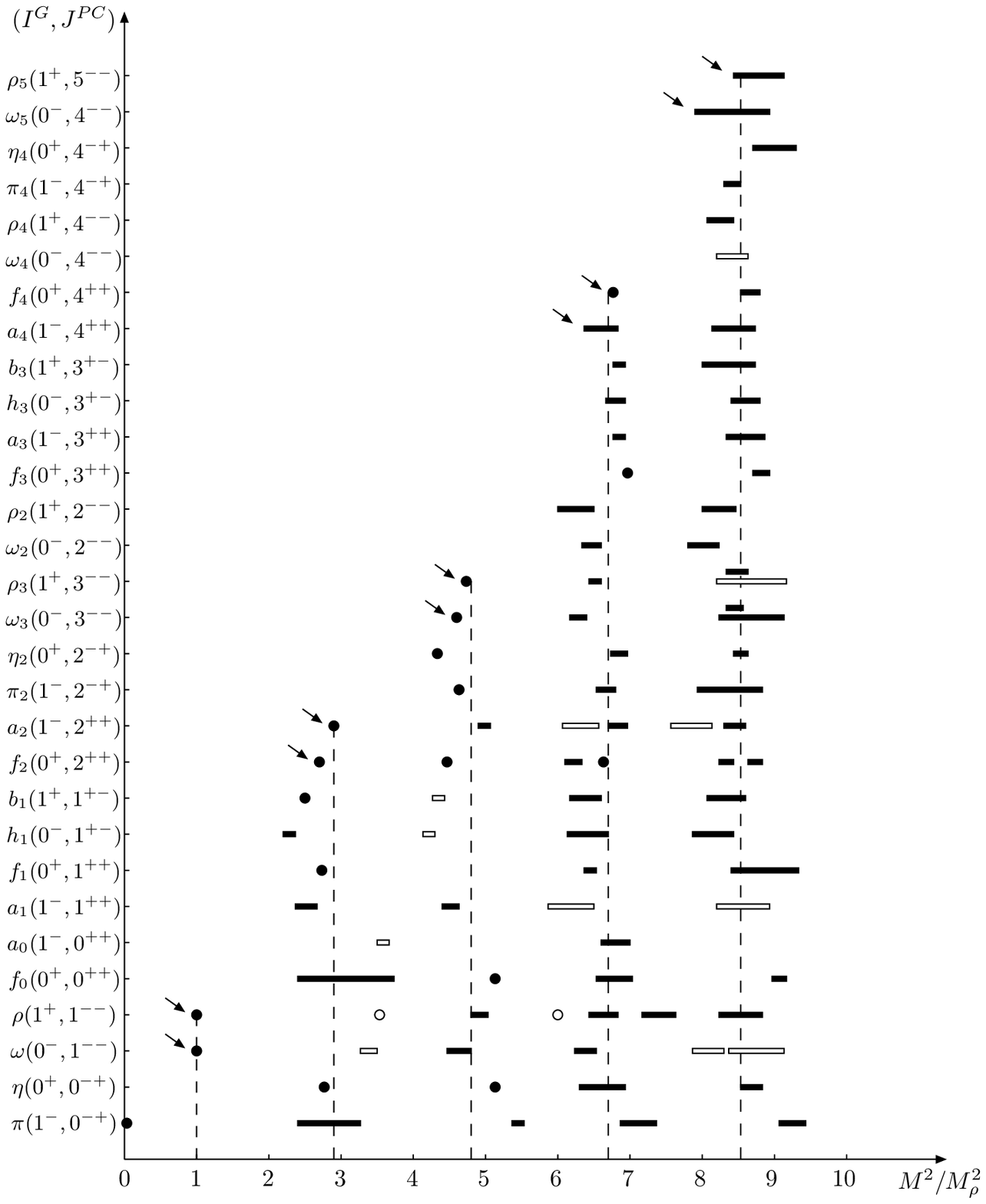}
\vspace{-8cm}
\caption{\small The spectrum of light non-strange mesons in units of $M_{\rho(770)}^2$.
The data is taken from Table~1.
Experimental errors are indicated.
Circles stay when errors are negligible. The dashed lines mark the
mean (mass)$^2$ in each cluster of states and the open strips and circles denote
the states marked as (?) in Table~2.
The arrows indicate the $J>0$ mesons which have no chiral partners
(the hypothetical chiral singlets, see the discussions in Sect.~4).}
\end{figure}
\end{center}

\vspace{-1.3cm}

Let us estimate quantitatively to what extent the outlined picture works. For this
purpose we perform the following procedure. First, we take the quadratic mean values
for the masses in each cell of Table~2. The result is given in Table~3. Second, assuming the
parametrization of spectrum dictated by Eq.~\eqref{mm},
\be
\label{mm2}
M^2(L,n)=AL+Bn+C,
\ee
we find the best fit for the plane $M^2(L,n)$ by means of the least-squares method
applied to two variables $(L,n)$, with
the values being taken from Table~3. The parametrization~\eqref{mm2} is not expected to
be operative for the $(0,0)$-states, thus, they are excluded from the fit. Keeping in mind
the discussions in the previous section, we distinct three cases: the whole spectrum
($\bar{M}^2$), the spectrum for the S-wave states ($\bar{M}_s^2$), and the one for the
states with $L>0$ ($\bar{M}_{ns}^2$). The results of our fit are presented in Table~4.

\begin{table}
\caption{\small Quadratic mean values in the $(L,n)$ blocks of Table 2 (in MeV).}
\vspace{0.2cm}
\begin{tabular}{cccccc}
\hline
\hline
\begin{tabular}{c}
\begin{picture}(15,15)
\put(0,12){\line(1,-1){15}}
\put(-2,-3){$L$}
\put(10,7){$n$}
\end{picture}\\
\end{tabular}
& 0 & 1 & 2 & 3 & 4 \\
\hline
0
&
\begin{tabular}{c}
---\\
\end{tabular}
&
\begin{tabular}{|c|}
\hline
$1373$\\
\hline
\end{tabular}
&
\begin{tabular}{c}
$1780$\\
\end{tabular}
&
\begin{tabular}{||c||}
\hline
$1995$\\
\hline
\end{tabular}
&
\begin{tabular}{c}
$2242$\\
\end{tabular}
\\
1
&
\begin{tabular}{|c|}
\hline
$1294$\\
\hline
\end{tabular}
&
\begin{tabular}{c}
$1668$\\
\end{tabular}
&
\begin{tabular}{||c||}
\hline
$1970$\\
\hline
\end{tabular}
&
\begin{tabular}{c}
$2256$\\
\end{tabular}
&\\
2
&
\begin{tabular}{c}
$1673$\\
\end{tabular}
&
\begin{tabular}{||c||}
\hline
$1980$\\
\hline
\end{tabular}
&
\begin{tabular}{c}
$2260$\\
\end{tabular}
&  &\\
3
&
\begin{tabular}{||c||}
\hline
$2024$\\
\hline
\end{tabular}
&
\begin{tabular}{c}
$2273$\\
\end{tabular}
&  &  &\\
4
&
\begin{tabular}{c}
$2266$\\
\end{tabular}
&  &  &  &\\
\hline
\hline
\end{tabular}
\end{table}

\begin{table}
\caption{\small The spectrum obtained by the least-squares method (in GeV$^2$).
The last column displays the predicted mass of the ground state (in GeV).}
\vspace{0.2cm}
\begin{tabular}{l|c|c|c}
\hline
\hline
\phantom{$\dfrac12$} & $AL+Bn+C$ & $a(L+bn+c)$ & $\sqrt{C}$ \\
\hline
$\bar{M}^2$\phantom{$\dfrac12$} & $1.103L+1.102n+0.686$ & $1.103(L+n+0.622)$
& 0.828\\
$\bar{M}_{ns}^2$\phantom{$\dfrac12$} & $1.178L+1.135n+0.473$ & $1.178(L+0.963n+0.402)$
& 0.688\\
$\bar{M}_s^2$\phantom{$\dfrac12$} & $1.023n+0.957$ & $1.023(n+0.935)$
& 0.978\\
\hline
\hline
\end{tabular}
\end{table}

Thus, our phenomenological analysis confirms the law
\be
\label{law}
M^2(L,n)\sim L+n
\ee
for the spectrum of mesons composed of the up and down quarks.
Of course, our result is obtained for a particular $(L,n)$-assignment of the states,
which is not unique. Indeed, as can be seen from Table~2 and/or Fig.~1, the assignment
of some S-wave states to a particular cluster is ambiguous, hence, they can be reshuffled
in various ways in the cells of Table~2. However, since the statistical weight of these
states is relatively small, this ambiguity does not change the result drastically.
The exclusion of one-star states does not affect the fit strongly as well.
In any case the value of parameter $b$ Eq.~\eqref{1} lies in the interval
\be
0.9<b<1.1,
\ee
with a rather high confidence level, {\it i.e.}, it is compatible with $b=1$ within
the experimental errors.

\section{Discussions}

The spectrum~\eqref{law} (more exactly, its relativistic form, Eq.~\eqref{mj})
was first obtained within the old dual amplitudes~\cite{LS,Sh,ven}. Later it was
realized that this spectrum is a feature of the relativistic Nambu-Goto open string
since the frequencies of classical oscillatory and rotational motions of this string
coincide at the same energy (see, {\it e.g.},~\cite{zw}).
Recently the spectral law~\eqref{law} was reported within the AdS/QCD approach~\cite{ads,fork,fork1},
in the latter reference a holographic dual of QCD was proposed
which gives the slope $\frac12$ just as the WKB method.
The existing quark models are unable to reproduce Eq.~\eqref{law},
as was emphasized recently in~\cite{bicudo}, a solution of this
problem is a challenge for future quark models. In a broad sense,
spectrum of any such model should depend on a single "principal quantum
number"~\eqref{pqn}. An attempt to relate the existence of this
quantum number to confinement realized through the area law for
Wilson loops was undertaken in~\cite{simonov}.

In the last column of Table~4 we give an extrapolation of obtained
mass spectra to the mass of $(0,0)$-mesons for $L=0$ and $L>0$
states. The experimental masses of $\rho(770)$ and $\omega(782)$
mesons lie in between. This could mean that the physical vector
ground states represent a strong mixture of both S-wave and D-wave
components. On the other hand, it may be that the deviation from
the prediction is caused mainly by the quark/antiquark spin-spin
interaction. Within the WKB method a mass of the ground state is directly
related to a value of constant $\gamma$ in Eq.~\eqref{3}. In this respect
it is interesting to note that the values of parameter $c$ in Table~4
are close to those of $\gamma$ given by the WKB approximation without account of
both the quark spin and the string rotation. Thus, we see that a simple
application of semiclassical methods yields the result which is rather close
to the averaged experimental one. It would be interesting to develop the subject in future.

Given the phenomenological fit, a question appears: Which effective string model
proposed in the literature is in the best agreement with the
experiment? We think that this is a model advocated
in~\cite{baker}. First of all, it reproduces Eq.~\eqref{mj}.
Second, for the mesons with $L=J+1$ this model (if one leaves the quantization condition
$J=L+\frac12$ exploited in~\cite{baker}) gives an intercept
close to the reality: $c=\frac{5}{12}\approx 0.417$.

The next issue we consider is the asymptotic chiral symmetry restoration
in the highly excited states. Namely, we would like to draw attention to the
following aspect of the problem: All states on the lowest leading Regge trajectories
(the $\rho$, $\omega$, $f_2$, and $a_2$ trajectories)
do not possess the chiral partners. In Fig.~1 these states reside on the top of the
depicted meson clusters and they are indicated by arrows.
One can make even a stronger statement:
In the case of states belonging to these trajectories there are no particles
with equal spin in a close mass region. This experimental fact
(yet not well established) contradicts to the usual effective
chiral symmetry restoration scenario (for a recent review see~\cite{glozman}) where each
light hadron has to have a chiral partner with close mass at high enough energies,
approximately since 1.7~GeV
(we remind that the chiral transformations relate particles with equal spins).
Meanwhile, the phenomenon is a natural consequence of string-like
nonrelativistic spectral formulas like Eq.~\eqref{law}:
Since the parity of quark-antiquark system is defined as
\be
\label{par}
P=(-1)^{L+1},
\ee
one of parity-conjugated
meson towers has a lower~$L$, hence, the ground state in this tower is sentenced to be a parity
bachelor. This mechanism naturally implements the chiral symmetry breaking in the hadron
spectrum. For instance, the ground $\rho$- and $a_1$-mesons cannot be degenerate because
they are $(0,0)$ and $(1,0)$ states correspondingly (in the notation $(L,n)$) and the sum
$L+n$ is different. This effect is repeated at larger~$L$ for the states
on the leading trajectories, up to the highest available mass range.

Thus, the absence of chiral partners for the leading meson trajectories seems to be an
experimental fact, nevertheless, is this a law of Nature? A possible answer is negative ---
just nobody in the past performed a systematic search for these partners. Say, some states
on the leading nucleon Regge trajectory
possess the parity doublers with a close mass. However, at present the absence of parity
doubling on the leading meson Regge trajectories is a rather definite experimental
result\footnote{I am grateful to D. V. Bugg for elucidating me this point.}.
If the chiral symmetry gets effectively restored since approximately 1.7~GeV, then the states
$\rho_3(1690)$ and $\omega_3(1670)$, $f_4(2050)$ and $a_4(2005)$,
$\rho_5(2300)$ and $\omega_5(2250)$ have to possess the chiral partners. The resonances
$\rho_3(1690)$ and $\omega_3(1670)$ are well known, but nobody observed the corresponding
parity partners, which ought to be very easy to find. Instead, the Crystal Barrel revealed
very conspicuous $a_3(2031)$ and $f_3(2048)$, which are the parity partners of
$\rho_3(1982)$ and $\omega_3(1945)$ correspondingly. There is no sign of anything near
the well seen resonances $f_4(2050)$ and $\rho_5(2300)$ in the Crystal Barrel.
Thus, the only experiment designed for the systematic search of the non-strange
meson resonances above 1.9~GeV did not discover the chiral partners for the states
on the leading trajectories. One should wait for another similar experiment to
confirm or refute the result.

As discussed above, the hadron string picture naturally leads to the existence of
chiral-singlet states if one accepts the standard definition of parity for the
quark-antiquark system, Eq.~\eqref{par}. A possible objection is that this definition
is nonrelativistic. It may be that if one redefines the parity somehow then the chiral-singlet
states will disappear from the hadron string spectrum and the latter will be compatible
with a usual chiral symmetry restoration scenario (for a recent attempt see~\cite{gloz2}).
Our conclusion concerning the experimental existence of chiral singlets, however, does not
depend on a concrete definition of parity: if there is no state with equal spin in a close
vicinity of some hadron, then this hadron is a chiral singlet. But what does "close" mean?
The cluster structure of mass distribution seen in Fig.~1 provides an answer to this
question: There is no state with equal spin in the given cluster, {\it i.e.}, the notion
"close" is defined with respect to the typical distance between two neighbouring clusters.

Finally, it should be emphasized that the existence of chiral singlets does not necessary
contradict to the effective chiral symmetry restoration in the highly excited states.
Say, the symmetry particle-antiparticle has singlets, {\it e.g.}, the $\pi^0$-meson.
The situation with the chiral symmetry could be somewhat similar.

\section{Conclusions}

We have reviewed briefly some recent attempts to understand a
broad degeneracy existing in the spectra of light non-strange
mesons, the X-symmetry. The degeneracy can be summarized compactly
as the spectral law $M^2(L,n)\sim L+n$, which we checked using the
latest experimental information and
the agreement is found to be remarkable. Needless to say that
this law, if true, is of great importance --- it explains the X-symmetry
and plays a powerful selective role both for the model building and for the
hadron systematics. For instance, it seems to be not compatible with a recently
proposed $\tilde{U}(12)_{SF}\times SO(3,1)_L$ classification scheme for mesons~\cite{ish},
which predicts the spectrum (see also~\cite{yam}) $M^2(L,n)\sim L+2n$.

A natural candidate for a dynamical model reproducing the expected mass spectrum and, hence,
generating the X-symmetry is argued to be an effective hadron string.
At present it is not clear if the hadron strings really exist as physical objects.
It may be that
looking at hadrons through the magnifying glass we would see something different. Nevertheless,
the strings are shown to emerge quite naturally as an organizing principle for hadron
spectroscopy. The concept of hadron strings combines the theoretical simplicity and explicative
power, in this respect it seems to be unrivalled presently,
although some interesting alternative approaches exist as well.

We demonstrated that a constant correction (the intercept) to the spectrum under consideration,
Eq.~\eqref{law}, can be understood quasiclassically, namely, within the WKB approximation. This supports
the idea that the highly excited states are, in essence, semiclassical systems.

The hypothetical X-symmetry seems to include the asymptotic chiral and axial symmetries as
a particular case. On the other hand, given the modern experimental data, the X-symmetry should
lead to the existence of chiral singlet states. A development of this subject
is a challenge for future models aiming at the description of
meson spectra.

\section*{Acknowledgments}
I am grateful to M. Shifman and A. Vainshtein for private communications and to the
Department of Structure and Constituents of the Matter (especially to Prof. D. Espriu)
of Barcelona University, where a part of the work was carried out.
The work was supported by CYT FPA, grant 2004-04582-C02-01, CIRIT GC, grant 2001SGR-00065,
RFBR, grant 05-02-17477, grant LSS-5538.2006.2, and by the
Ministry of Education of Russian Federation, grant RNP.2.1.1.1112.

\end{document}